\documentclass[twocolumn]{jpsj3}
\usepackage{txfonts}
\usepackage{bm}

\title{Snapshot spectra in the world-line quantum Monte Carlo for one-dimensional quantum spin systems}
\author{Kouichi Seki, and Kouichi Okunishi}
\inst{Department of Physics, Niigata University, Niigata 950-2181, Japan}

\abst{
We introduce a snapshot density matrix and snapshot spectrum for world-line (WL) quantum Monte Carlo simulations, by integrating out the continuous-imaginary-time index of WL snapshots.
For the transverse-field Ising chain, we reveal fundamental properties of the snapshot spectrum as follows.
(i) In the ordered phase, the isolated peak corresponding to the maximum eigenvalue appears, indicating the classical nature of WL configurations.
(ii) In the disordered phase, the spectrum converges to the dome-like distribution characterized by an effective aspect ratio of the snapshots.
(iii) At the critical point, the power-law distribution is observed with an appropriate scaling analysis of the fixed aspect ratio.
For the XXZ chain, we also obtain numerical results consistent with the above properties.
In addition, the representation-basis dependence of the snapshot spectrum is also discussed.
}

\begin{document}
\maketitle

\section{Introduction}
Analyses of entanglements for quantum many-body systems often provide essential information that cannot be accessed by the usual bulk physical quantities.
For instance, the area law of the entanglement entropy and the non-trivial topological entropy have been intensively investigated to clarify a variety of novel properties of quantum many-body systems\cite{RevModPhys.82.277, RevModPhys.80.517, calabrese2004entanglement, kitaev-preskill}.
Moreover, entanglement structures embedded in quantum many-body systems have been providing a guiding principle for designing quantum simulation methods such as density matrix renormalization group\cite{dmrg} and tensor network simulations\cite{CTMRG, TPS, TRG, mera, PEPS, tensornet}.
On the other hand, quantum Monte Carlo (QMC) is also a quasi-exact numerical simulation method, which enables us to efficiently estimate expectation values of bulk physical quantities and correlation functions at finite temperature.
In general, however, the QMC is not suitable for analysis of the entanglement, except for the $n = 2$ R{\'e}nyi entropy where the valence-bond-solid basis is available\cite{PhysRevB.82.224414}.
Thus, it may be an intriguing problem to investigate how information analogous to the quantum entanglement can be extracted from QMC simulations.

For two-dimensional (2D) classical spin systems, very recently, a similar but different idea of the quantum entanglement was proposed for classical MC; a snapshot spectrum and snapshot entropy were defined for the singular value spectrum of a snapshot generated by a MC simulation\cite{matsueda2012holographic, imura2014snapshot, matsueda2015proper}.
It was shown that the snapshot spectrum gave hierarchical decomposition of snapshot images in Ref. [\citen{matsueda2012holographic}], and a characteristic behavior of a snapshot entropy was also observed.
In Ref. [\citen{imura2014snapshot}], further, it was clarified that the distribution function of the snapshot spectrum successfully captures intrinsic features of phase transitions; the distribution function in the high-temperature limit is described by the random matrix theory, while the low-temperature phase is characterized by zero-eigenvalue condensation of the spectrum.
At the critical point, moreover, the distribution function exhibits a power-law decay with a novel critical exponent that can be attributed to a correlation function matrix.

Since the 2D classical system corresponds to the 1D quantum system through the Suzuki-Trotter decomposition, it is basically expected that the above analysis for the 2D classical systems is applicable to the 1D quantum systems\cite{trotter1959product, suzuki1976relationship}.
Nevertheless, in the QMC of loop-type algorithm, which is one of the most popular simulation approaches for quantum spin systems, snapshots of spin configurations are represented as continuous and winding world-lines (WLs)\cite{evertz1993cluster, Kawashima_loopspin, beard1997simulations, PhysRevE66046701, kawashima2004recent}, which are apparently different from spin configurations on a 2D lattice in the classical MC.
Thus, we need an appropriate definition of the snapshot matrix to directly discuss the snapshot spectrum representing weight for hierarchical decomposition of the WLs.

In this paper, we introduce snapshot spectra for WL configurations generated by QMC simulations, and then investigate their distribution functions for the transverse-field Ising chain and the $S = 1/2$ XXZ chain in details.
In particular, we analyze how quantum fluctuation affects the snapshot spectra and discuss its relation to features of typical WL configurations.
We also clarify how quantum critical phenomena can be represented by the snapshot spectrum in the low-temperature limit, where the relation with the correlation-function matrix is important as in the case of 2D classical systems.
In addition, we mention the dependence of the snapshot spectrum on the representation basis of WLs.

This paper is organized as follows.
In the next section, we introduce the snapshot density matrix (SDM) for WL snapshots generated by the loop algorithm of QMC, by integrating out the imaginary-time index.
In Sec. III, we investigate snapshot spectra for the transverse-field Ising chain in details to clarify their fundamental properties.
In Sec. IV, we also analyze snapshot spectra in the $S^x$-diagonal-basis representation of WLs.
In Sec. V, snapshot spectra for the XXZ chain are analyzed in details.
In Sec. VI, we summarize our results.

\section{Snapshot spectrum for WLs}
In order to analyze snapshot spectra for quantum spin systems, a straightforward approach is to map the quantum systems into the corresponding classical lattice systems through the Suzuki-Trotter decomposition\cite{trotter1959product, suzuki1976relationship}.
However, a careful extrapolation with respect to the Trotter number is needed to restore the continuous-imaginary-time limit relevant to the original quantum spin systems.
In addition, QMC simulations widely used for recent researches of quantum spin systems are based on loop-type algorithms for WLs of spins/particles\cite{beard1997simulations,PhysRevE66046701,kawashima2004recent}, where the continuous-imaginary-time limit is automatically taken into account.
In this sense, it could be convenient to directly deal with the snapshot spectrum for WLs with the continuous imaginary time.
In this section, we explain the snapshot matrix defined on the 2D classical lattice based on discretized WL configurations, and then consider its continuous-imaginary-time limit.

As an example, let us consider the transverse-field Ising chain, which is written as
\begin{equation}
\mathcal{H} = - J\sum_n^{L} \hat{S}_n^z\hat{S}_{n+1}^z - \Gamma \sum_n^{L} \hat{S}_n^x,\label{TVI}
\end{equation}
where $\hat{S}_n^{x,y,z} $ are $S = 1/2$ spin operators at the $n$th site, $J$ denotes the exchange coupling, and $\Gamma$ represents a transverse magnetic field.
Also, $L$ is a number of spin sites in the real-space direction and the boundary condition of the chain is assumed to be periodic.
In the following, we explain the SDM and snapshot spectrum for the transverse-field Ising chain~(\ref{TVI}), where we can invoke results of the 2D Ising model.
However, generalization to other quantum spin systems is straightforward.

\begin{figure}[tb]
\centering\includegraphics[width=\linewidth]{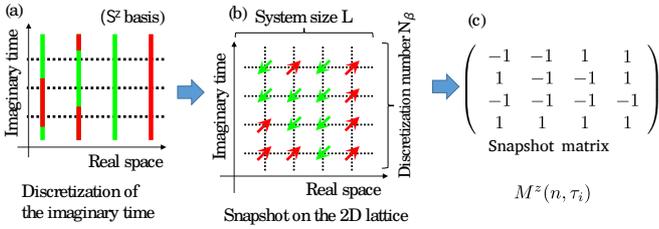}
\caption{
(Color online)
(a) Schematic diagram of a continuous WL configuration for the transverse-field Ising chain by a WL-QMC ($S^z$ basis representation).
The red and green lines respectively show the up and down spins.
(b) Spin configuration on the 2D lattice with the discretized imaginary-time index.
(c) Snapshot matrix $M^z(n, \tau_i)$ for the spin configuration of (b), where the real-space  and imaginary-time directions  respectively correspond to the column and row indices of $M^z$.
}\label{WLtoMatrix}
\end{figure}

Suppose that a WL configuration of spins in the $S^z$-diagonal basis at the equilibrium of an inverse temperature $\beta$ is obtained by a QMC simulation.
Note that the relation between matrix elements in the Hamiltonian and structure of WLs generated by the loop-type QMC algorithm are briefly summarized in Appendix \ref{appendix0}.
We then write this WL configuration at a site $n$ and imaginary time $\tau$ as $S(n, \tau)$, and further define a snapshot matrix as $M^z(n, \tau) \equiv 2 S(n, \tau)$, where $M^z(n, \tau)$ takes $\pm 1$, like the 2D classical Ising model in Refs.[\citen{matsueda2012holographic, imura2014snapshot}].
Of course, $\tau$ is a continuous index of the imaginary time with the period $\beta$, so that $M^z(n, \tau)$ can be viewed as a highly anisotropic matrix.
In analogy with a snapshot of the 2D classical case, thus, we discretize the imaginary time, $\tau_i = \beta/N_\beta \times i$ with $i =1\cdots N_\beta$, where $N_\beta$ corresponds to the Trotter number (see Fig. \ref{WLtoMatrix}).
A useful point on the discretized snapshot matrix $M^z(n, \tau_i)$ is that it basically has the same structure as that of the 2D Ising spins on the $L \times N_\beta$ lattice.
Then, we can analyze the $N_\beta$ dependence of the snapshot spectrum and recover the original WL configurations by taking the $N_\beta \to \infty$ limit again.

As in the case of the 2D classical Ising model, we consider singular-value decomposition (SVD) of the snapshot matrix,
\begin{equation}
M^z(n, \tau_i) = \sum_{l=1}^{{\rm min}(L, N_\beta)} U_l(n)\sqrt{\omega_l} V_l(\tau_i),
\label{svd}
\end{equation}
where $\sqrt{\omega_l}$ is singular value arranged in the decreasing order and $U_l$ ($V_l$) is the corresponding left (right) singular vector.
Note that the number of singular values is bounded by ${\rm min}\{L, N_\beta\}$, and the singular vectors satisfy the orthonormal relation, $^{\rm t}U_lU_{l'} = {}^{\rm t}V_{l}V_{l'} = \delta_{l, l'}$.
We then define the SDM by contracting the imaginary-time index $\tau_i$ as
\begin{align}
\rho^z(n, m) &= \frac{1}{N_\beta}\sum_{i=1}^{N_\beta} M^z(n, \tau_i)M^z(m, \tau_i) \nonumber \\
&= \frac{1}{N_\beta}\sum_{l=1}^{{\rm min}(L, N_\beta)} U_l(n)\omega_l U_l(m),
\label{SDMr}
\end{align}
where the eigenvalue is written as square of the singular value of $\sqrt{\omega_l}$.
The normalization of the SDM is ${\rm Tr}\rho^z = L$, since the diagonal elements of $\rho^z$ are unity for the Ising spin.
Note that another SDM by contracting the lattice index $n$ gives the same eigenvalues $\omega_l$ as Eq. (\ref{SDMr}).

We diagonalize the SDM $\rho^z$ for a single snapshot of WLs generated by a QMC simulation to have the snapshot spectrum $\left\{\omega_l\right\}$.
What we focus on is that $\omega_l$ extracts the hierarchical structure embedded in a snapshot matrix.
As shown in Appendix \ref{appendix1}, the largest eigenvalue $\omega_1$ basically represents the global structure of WLs.
The smaller $\omega_l$ becomes, the shorter range structures of the WL snapshot are systematically reproduced.
Thus, we can expect that the distribution of $\omega_l$ may provide a novel characterization of the nature of the system, different from bulk physical quantities.
Nevertheless, a single snapshot for a finite-size system generated by a QMC simulation contains large statistical fluctuations.
We thus define the distribution function $P(\omega)$ as
\begin{equation}
P(\omega) = \frac{1}{L}\left\langle\sum_l\delta(\omega - \omega_l)\right\rangle_{\rm MC},
\label{paverage}
\end{equation}
where $\langle \cdots \rangle_{\rm MC}$ represents the sample average in a QMC simulation.
In practical situations, we calculate $P(\omega)$ as a histogram of the eigenvalues with a certain discretization of $\omega$, during a run of QMC simulation.
We then discuss features of this distribution function for various quantum spin chains.

Although the above definition of the SDM has a clear correspondence to that of the 2D classical system, the resulting snapshot spectrum includes an error due to discretization in the imaginary-time direction.
We discuss the continuous-imaginary-time limit, i.e. $L\ll N_\beta \to \infty$, for the snapshot spectrum.
As $N_\beta$ increases, the interval $\Delta \tau = \beta/N_\beta$ of the imaginary-time direction becomes smaller than intervals of scattering points/kinks of winding WLs, and then adjacent spin configurations between $M^z(n, \tau_i)$ and $M^z(n, \tau_{i+1})$ finally become indistinguishable.
This implies that the rank of the snapshot matrix is eventually bounded at the order of $N^* \sim \beta/\Delta \tau_s$, where $\Delta \tau_s$ denotes the average interval of scattering points/kinks in winding WLs.
Thus, we can expect that the SDM also converges in the $N_\beta \to \infty$ limit, where the sum with respect to $i$ in Eq. (\ref{SDMr}) turns out to be the integral of the continuous imaginary time $\tau$.
Taking account of $d\tau \simeq \beta/N_\beta$, we explicitly have
\begin{equation}
\rho^z(n, m) = \frac{1}{\beta}\int_0^{\beta}M^z(n, \tau)M^z(m, \tau)d\tau,
\label{SDMc}
\end{equation}
where the normalization is ${\rm Tr} \rho^z = L$.
Here, it should be noted that, although $M^z(n, \tau)$ has a continuous index $\tau$, the dimension of Eq. (\ref{SDMc}) is $L \times L$, where the number of the eigenvalues is fixed at $L$.
We have actually investigated the $N_\beta$ dependence of the snapshot spectrum, and then confirmed that numerical results converge to those of Eq. (\ref{SDMc}), as $N_\beta$ increases.
In the following, we thus adopt Eq. (\ref{SDMc}) as the definition of the SDM for continuous-imaginary-time WLs.

\section{Transverse-field Ising chain}
In this section, we consider the snapshot spectrum for the transverse-field Ising chain.
WL configurations in the $S^z$-diagonal basis are generated by QMC simulations based on the loop algorithm\cite{kawashima2004recent}.
In practical QMC simulations, we normalize the exchange coupling with $J=1$ for simplicity.
The discretization width for histograms of $P(\omega)$ is $\Delta \omega = \epsilon L$ with $\epsilon = 1.0 \times 10^{-4}$ -- $1.2 \times 10^{-2}$, and typical number of samples is $1.0 \times 10^4$ -- $1.6 \times 10^5$.

In the high-temperature limit, the snapshot spectrum becomes trivial independently of the groundstate phases, since the WL configurations for $T \gg J$ are represented as straight lines with no kink reflecting the classical behavior of spins.
This is because the length of WLs for $\beta J \ll 1$ is very short, compared with the energy scale of the quantum effect.
Therefore, the imaginary-time index $\tau$ can be suppressed and the snapshot matrix $M^z(n, \tau)$ can be eventually reduced to be the column vector whose elements are  $+1$ or $-1$, for which  the matrix rank of $\rho^z(n, m)$ given by Eq. (\ref{SDMc}) is unity.
Then, the SDM trivially has the maximum eigenvalue $\omega_1 = L$ and the other eigenvalues degenerate at $\omega=0$, implying that $P(\omega)$ has two $\delta$-function-like peaks at $\omega = L$ and $0$.
Note that this behavior is also independent of the system size.

As temperature decreases, a lot of kinks due to the quantum fluctuation appear in WLs along the imaginary-time direction.
Accordingly, $P(\omega)$ changes from the $\delta$-function shape to the broadened peaks, reflecting the groundstate properties of the system.
Here, we briefly summarize the groundstates of the transverse-field Ising chain\cite{pfeuty1970}.
For $\Gamma/J < 0.5$, the groundstate is in the ordered phase, where spins are uniformly aligned in the classical ($S^z$) direction.
For $\Gamma/J > 0.5$, the groundstate is in the disorder phase, where spins randomly behave due to the significant quantum effect.
At $\Gamma/J=0.5$, the groundstate is critical and its universality belongs to the 2D Ising class.

\subsection{Ordered phase}
\begin{figure}[tb]
\centering\includegraphics[width=0.8\linewidth]{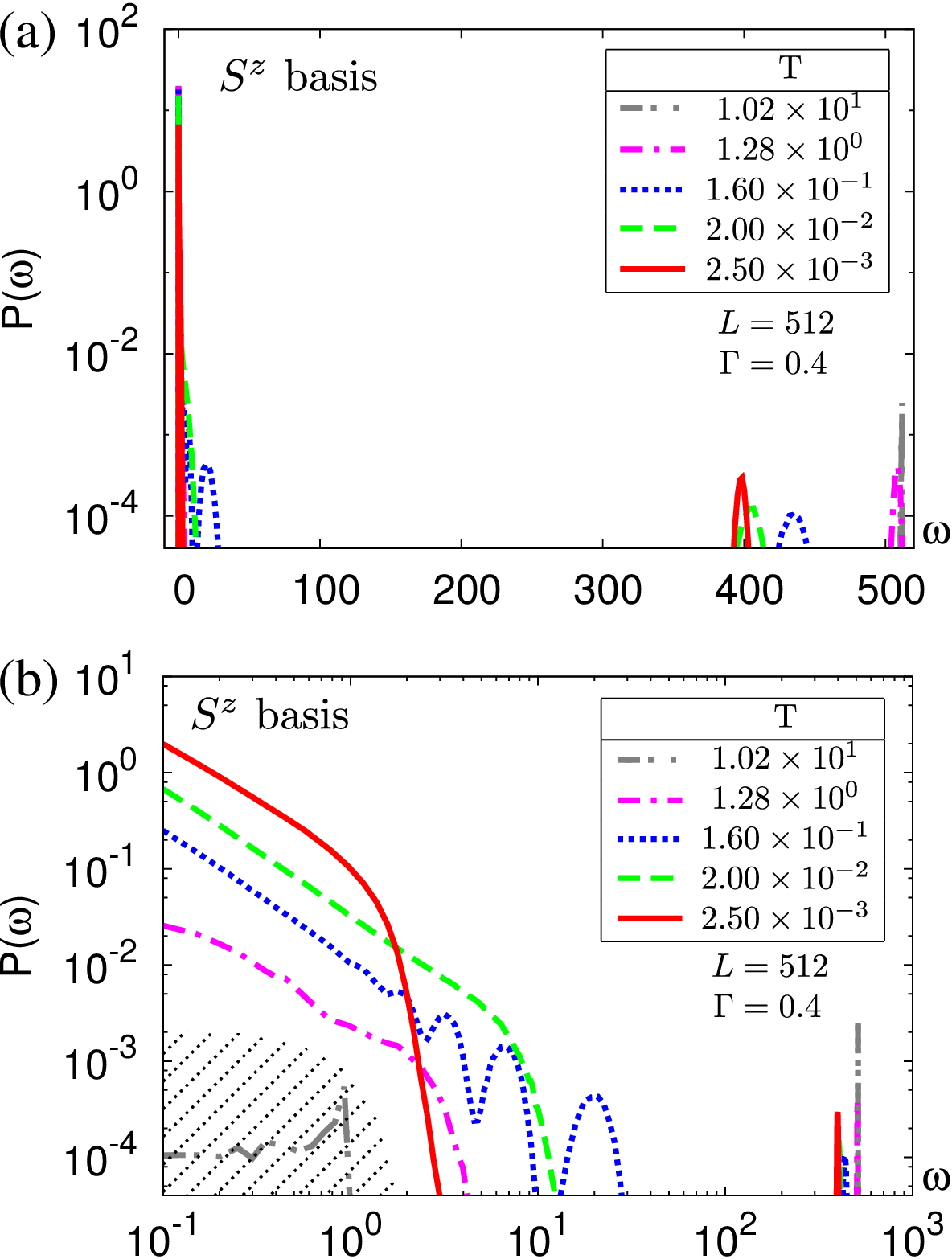}
\caption{
(Color online)
Temperature dependence of the eigenvalue distribution $P(\omega)$ for the transverse-field Ising chain in the ordered phase: $\Gamma = 0.4$ and $L=512$.
(a) Semi-logarithmic plot of $P(\omega)$, where two peaks emerge at $\omega = 0 $ and $\omega = L $ in the high-temperature limit ($T = 1.02\times 10^1$).
(b) Logarithmic plot of $P(\omega)$, where the tallest peak at $\omega=0$ is out of range.
The statistical error is negligible except for the shaded region.
}\label{TVI_Sz_order_H040_EVD}
\end{figure}

In Fig. \ref{TVI_Sz_order_H040_EVD}, we show temperature dependence of $P(\omega)$ for $\Gamma=0.4$ with $L=512$.
As described above, WL snapshots in the high-temperature region ($T=1.02\times 10^1$) are represented as straight lines and then, the snapshot spectrum has two peaks at $\omega=0$ and the maximum eigenvalue $\omega \simeq L(=512)$, which can be seen in the semi-log plot of Fig.  \ref{TVI_Sz_order_H040_EVD}(a).
We have numerically confirmed that the area of the peak at $\omega \simeq L$ is $1/L$ within the numerical accuracy, implying that this peak solely consists of the maximum eigenvalue of the SDM.

As temperature decreases, the quantum fluctuation due to the transverse magnetic field disturbs the classical situation of WLs.
In particular, a part of eigenvalues macroscopically degenerating at $\omega=0$ are lifted by the quantum effect, as can be seen in the log-log plot of Fig. \ref{TVI_Sz_order_H040_EVD}(b).
We then find that the crossover between the high-temperature and low-temperature behaviors occurs;
Around the crossover temperature scale $T\sim \Gamma/J=0.4$, $P(\omega)$ exhibits the multi-peak structure (See $T=1.60\times 10^{-1}$ in Fig. \ref{TVI_Sz_order_H040_EVD}(b)).
Note that these multi-peaks are confirmed to originate from a few large eigenvalues in the snapshot spectrum.
Below $T\sim \Gamma/J=0.4$, these peaks merge again into a continuous distribution in the low-$\omega$ region, while the isolated peak of the maximum eigenvalue is still maintained around $\omega \simeq 400$ even in the low-temperature limit.
This robust maximum eigenvalue originates from the fact that dominant number of WLs are stably aligned in the direction of the classical spin order against the weak quantum fluctuation.
Also, we have confirmed that 33\% of the eigenvalues still degenerate at $\omega=0$, reflecting the reduction of the matrix rank due to the classical order.
Thus, it can be concluded that the isolated peak of the maximum eigenvalue indicates the classical order of the groundstate.

\subsection{Disordered phase}\label{TVI_Sz_disorder}
\begin{figure}[tb]
\centering\includegraphics[width=0.8\linewidth]{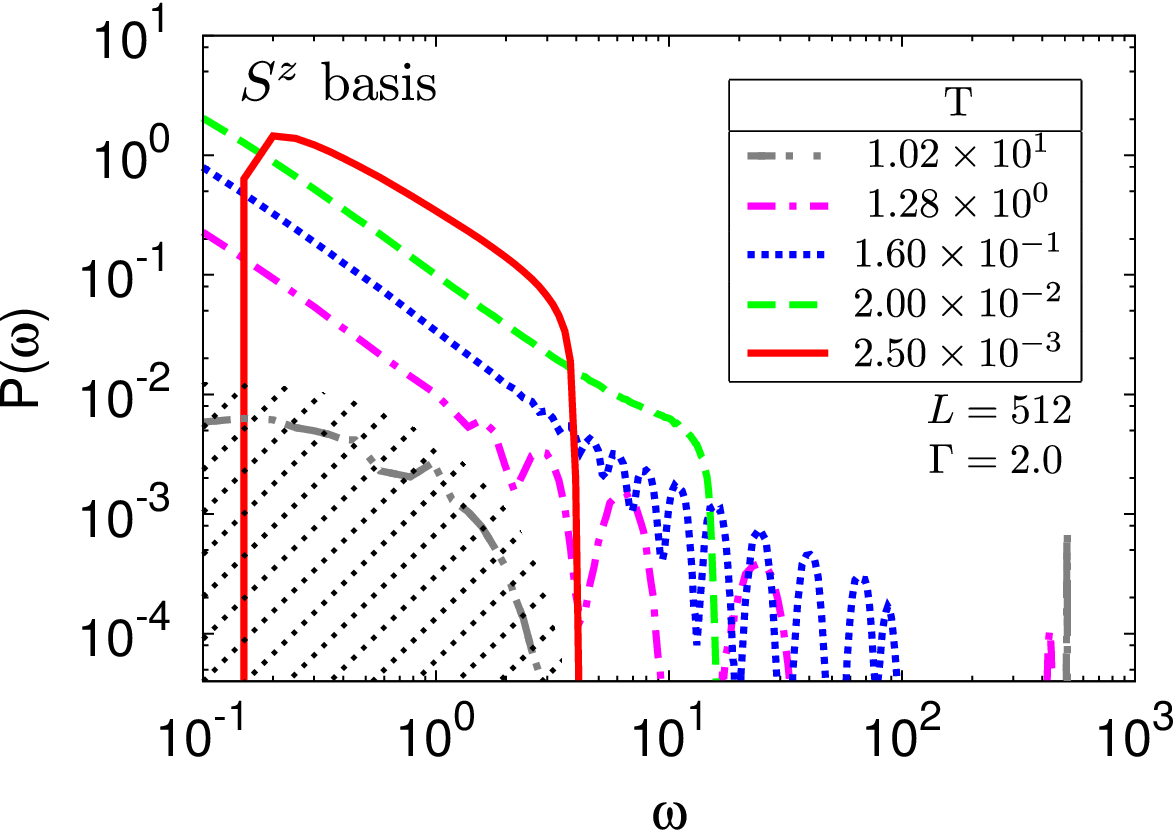}
\caption{
(Color online)
Temperature dependence of the eigenvalue distribution $P(\omega)$ for the transverse-field Ising chain in the disordered phase: $\Gamma = 2.0$ and $L=512$.
The statistical error is negligible except for the shaded region.
}\label{TVI_Sz_disorder_H200_EVD}
\end{figure}

In the disordered phase, the effect of the quantum fluctuation in the low-temperature region becomes significant.
The transverse-field term generates many kinks in WLs and destroys the classical order even at the groundstate, which yields nontrivial distribution of the snapshot spectrum.
In Fig. \ref{TVI_Sz_disorder_H200_EVD}, we show temperature dependence of $P(\omega)$ for $\Gamma=2.0$ and $L=512$.
At high temperature ($T=1.02\times 10^1$), the peak of the maximum eigenvalue solely emerges at $\omega \simeq 500$.
In the crossover temperature scale of $T\sim \Gamma/J(=2.0)$, the multi-peak structures corresponding to a few larger eigenvalues also appear in $\omega < 100$.
As temperature decreases further, however, these peak structures including the maximum-eigenvalue peak are smeared out and turn out to be a smooth single distribution.
At the lowest temperature $T=2.50\times10^{-3}$, moreover, we can see that $P(\omega)$ exhibits a doom-like shape around $\omega \simeq 1$ and the degeneracy at $\omega=0$ disappears.

In the low-temperature region, WL configurations basically reflect the properties of the groundstate, suggesting that $P(\omega)$ would converge to a certain distribution characterized by the groundstate spin configuration.
In order to analyze this low-temperature distribution, we remark two important parameters.
The competition between the quantum fluctuation and the spin-spin interaction is controlled by $\Gamma/J$, which plays the role of temperature in the corresponding 2D classical Ising model.
If $\Gamma/J$ is large enough in the disordered phase, WL configurations are eventually random in the spatial direction, so that a universal curve of the eigenvalue distribution can be expected, as in the case of the high-temperature limit of the 2D classical Ising model\cite{imura2014snapshot}.
Another parameter is the effective aspect ratio of the WL snapshot defined as
\begin{equation}
Q \equiv \frac{\beta \Gamma}{L},
\label{aspectratioQ}
\end{equation}
where $\beta \Gamma$ represents the effective system size in the imaginary-time direction.
Here, note that $\Gamma$ determines density of kinks embedded in WLs.
This effective aspect ratio $Q$ basically controls the shape of the low-temperature distribution of the snapshot spectrum.

\begin{figure}[tb]
\centering\includegraphics[width=0.8\linewidth]{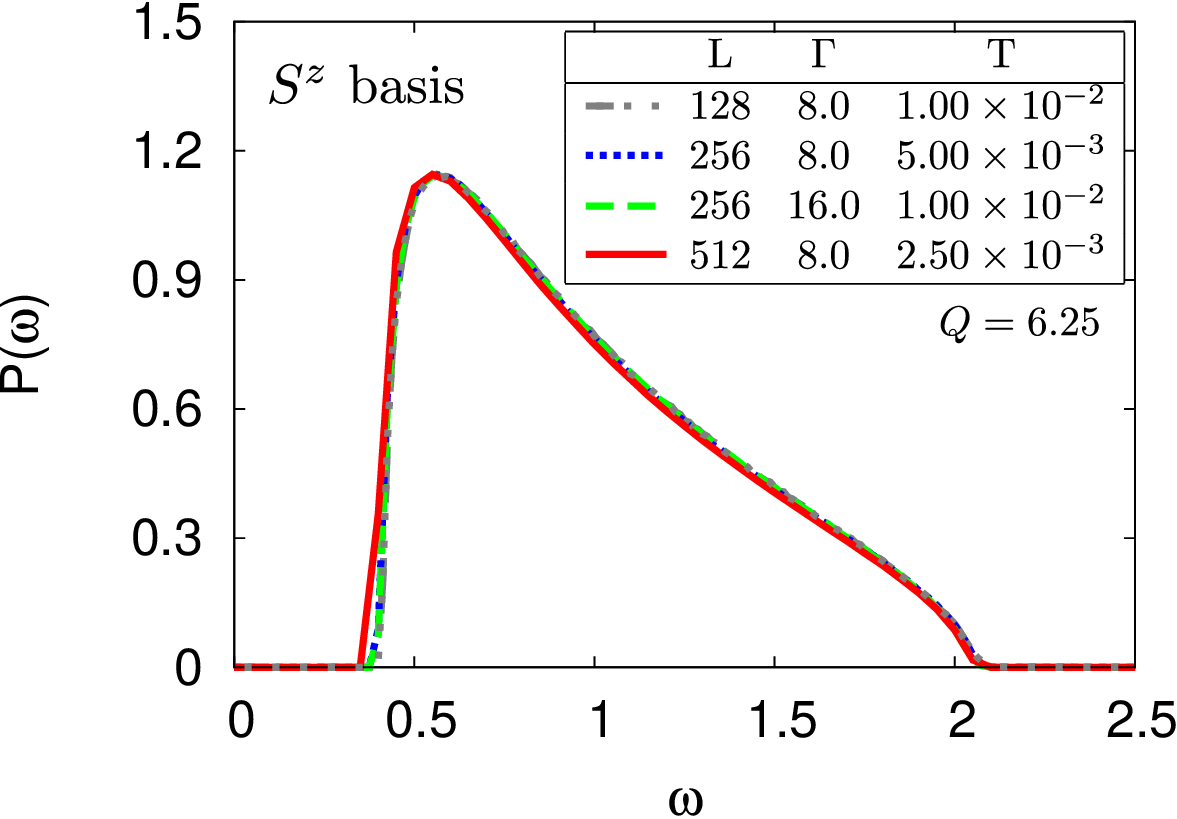}
\caption{
(Color online)
Low-temperature eigenvalue distributions $P(\omega)$ for sufficiently large $\Gamma$.
The curves for various $T$, $L$ and $\Gamma$ with $Q=6.25$ collapse onto the curve characterized by the effective aspect ratio $Q$.
}\label{TVI_Sz_disorder_Q625_EVD}
\end{figure}

In Fig. \ref{TVI_Sz_disorder_Q625_EVD}, we plot $P(\omega)$ for various $T$, $L$ and $\Gamma$ with a fixed aspect ratio $Q=6.25$.
In the figure, we can find that, for the sufficiently large $\Gamma/J(>8.0)$, $P(\omega)$ actually collapses into the universal curve characterized by $Q$.
In contrast to the 2D classical Ising model, however, this distribution for large $\Gamma/J$ is {\it not} described by the Wishart distribution based on the random matrix theory.
The continuous WLs of the transverse-field Ising chain always involve the auto correlation in the imaginary-time direction, though the correlation effect in the spatial direction is negligible for large $\Gamma/J$.
Meanwhile, the spin correlation for the 2D classical Ising model in the high-temperature limit can be negligible in the both spatial directions, to which the random matrix theory is applicable.

As $\Gamma/J$ approaches the critical point, the distribution is gradually extended to the large-$\omega$ region, because of the spatial correlation effect.
At the critical point, the distribution exhibits a power-law decay for $\omega\gg 1$, which is discussed in the next subsection in details.

\subsection{Critical point ($\Gamma = 0.5$)}
\begin{figure}[tb]
\centering\includegraphics[width=0.8\linewidth]{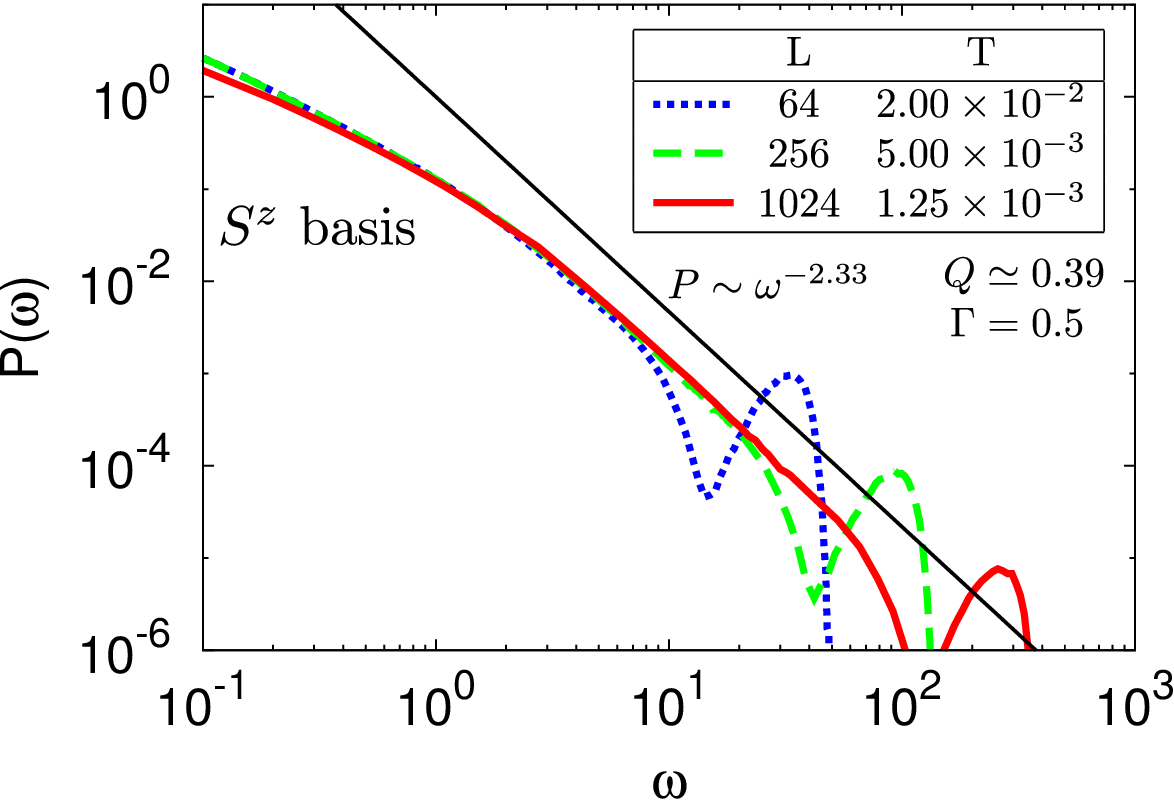}
\caption{
(Color online)
Log-log plot of the eigenvalue distribution $P(\omega)$ for the transverse-field Ising chain at the critical point ($\Gamma=0.5$), where the aspect ratio is fixed at $Q = 25/64 \simeq 0.39$.
The solid line indicates a slope with the exponent $\alpha=2.33$ for comparison.
}\label{TVI_Sz_critical_H0500_EVD}
\end{figure}

The transverse-field Ising chain exhibits the quantum critical behavior at $\Gamma/J=0.5$, where the highly fluctuating behavior of WLs can be observed in snapshots.
In accordance with the result of the 2D classical Ising model\cite{imura2014snapshot}, we assume that the SDM at the critical point can be associated with the correlation function matrix
\begin{equation}
\rho^z(n, m)\sim \langle S_n^z S_m^z \rangle \sim {\rm const} \times |n-m|^{-\eta},
\label{sdm_z}
\end{equation}
motivated by the self-averaging effect.
Note that the exponent $\eta =1/4$ for the transverse-field Ising model.
We can then diagonalize Eq. (\ref{sdm_z}) by Fourier transformation to have the dispersion relation
\begin{equation}
\omega(k) \sim \int_0^\infty r^{-\eta} e^{ikr} dr \, ,
\label{disp_omega}
\end{equation}
for $L\gg 1$.
If $\eta<1$, which is actually the case for the present $S^z$ basis, the integration of Eq. (\ref{disp_omega}) converges to $k^{\eta-1}$, which leads us to the power-law distribution, $P(\omega) \sim \omega^{-\alpha}$ with
\begin{equation}
\alpha\equiv \frac{2-\eta}{1-\eta}\, .
\label{exponent_alpha}
\end{equation}
As mentioned in the previous subsection, we should fix the effective aspect ratio of the snapshots for a systematical analysis.
In Fig. \ref{TVI_Sz_critical_H0500_EVD}, we show temperature/size dependences of $P(\omega)$ for $J=1$ and $\Gamma=0.5$ with the fixed $Q = 25/64 \simeq 0.39$.
In this figure, we can see humps near the distribution edges in the large-$\omega$ side, which are attributed to the peak of the maximum eigenvalue.
As $\beta$ increases, however, the height of the hump lowers with shifting its position to the larger-$\omega$ side, suggesting that the humps also follow a finite-size scaling.
Then, we can verify that the power-law region extends to the larger-$\omega$ side, as $\beta$ increases.
The exponent of the power-law distribution is estimated as $\alpha=2.33$, which is consistent with the solid line in Fig. \ref{TVI_Sz_critical_H0500_EVD} corresponding to $\alpha$ of the 2D Ising universality.
In this sense, we conclude that the snapshot spectrum of WL QMC is also capable of the proper quantum critical behavior by the appropriate scaling with the fixed aspect ratio $Q$.

Here, we should remark that the relation (\ref{sdm_z}) does {\it not} imply that the snapshot spectrum is always equivalent to the spectrum of the correlation function matrix;
In general, the SDM and correlation function matrix yield different spectra away from the critical point. 
An essential point is that the sample average of Eq. (\ref{paverage}) is taken after diagonalizing the SDM, whereas, for Eq. (\ref{disp_omega}), the correlation function matrix is diagonalized after the sampling average is taken.
In the high-temperature limit, for example, the snapshot spectrum has double peaks at $\omega =0 $ and $L$, while the spectrum of the correlation function matrix, which has only diagonal elements, has a single peak at $\omega\sim 1/4$.
Our observation of the spectra suggests that the order of the diagonalization and the sample average is irrelevant only at the critical point, which is an interesting numerical finding in this paper.
However, the mechanism of such a nontrivial correspondence at the critical point is an important future issue.

\section{Snapshot spectrum in $S^x$-diagonal basis}
The QMC simulations in the previous section were performed in the $S^z$-diagonal basis of spins, while a QMC simulation is also possible in the $S^x$-diagonal basis.
These two basis are related through the unitary spin rotation in the Hamiltonian level, where expectation values of observables are of course identical in the both representations of spins.
In the snapshot level, however, appearance of WL configurations in the $S^x$ basis is quite different from that in the $S^z$ basis.
In the $S^z$ basis, the transverse-field term locally flips spins, implying that the WLs change their spin labels along the imaginary-time direction.
In the $S^x$ basis, on the other hand, the $S^zS^z$ interaction term generates pair creation/annihilation or exchange of WLs, so that the WLs always draw closed loops, while the transverse-field term basically plays a role of the chemical potential controlling the density of loops of up/down spins.
In this section, we investigate how the snapshot spectrum of the transverse-field Ising chain is described in the $S^x$ basis.

Before we present numerical results, here, we comment on the technical point that the relaxation to the equilibrium in QMC simulations with the $S^x$ basis generally becomes worse, as $\Gamma$ increases.
In our case, we could not achieve proper relaxation for $\Gamma > 2.0$.
However, we can extract essential properties of the snapshot spectra within $\Gamma\le 2.0$.

\begin{figure}[tb]
\centering\includegraphics[width=0.8\linewidth]{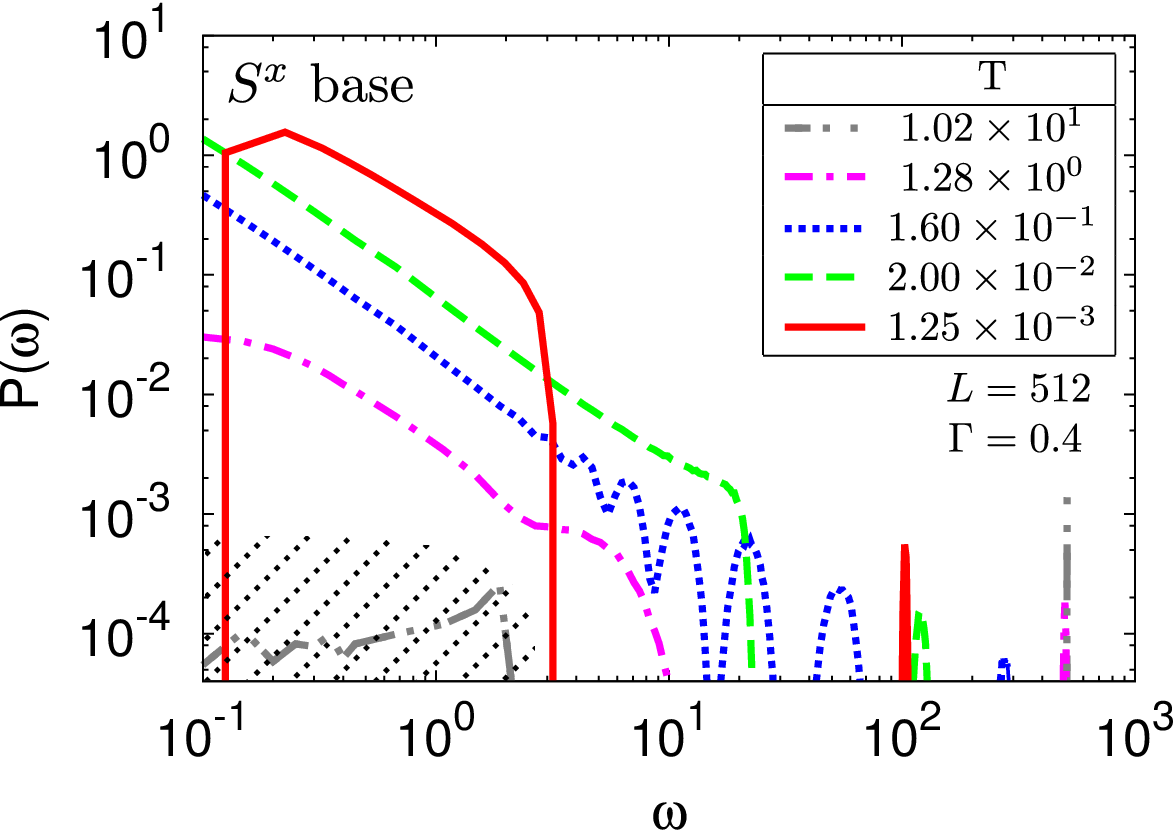}
\caption{
(Color online)
Temperature dependence of the eigenvalue distribution $P(\omega)$ in the $S^x$ basis for the transverse-field Ising chain: $\Gamma=0.4$ and $L=512$.
The statistical error is negligible except for the shaded region.
}\label{TVI_Sx_order_H040_EVD}
\end{figure}

In Fig. \ref{TVI_Sx_order_H040_EVD}, we first show the eigenvalue distribution in the ordered phase ($\Gamma=0.4$).
In the high-temperature limit where $\beta J$ is very small, the WLs are straight with no kink, and yield two $\delta$-function peaks at $\omega=0$ and $L$, as in the case of the $S^z$-diagonal representation.
Around the crossover temperature scale $T\sim J$, the multi-peak structures for a few largest eigenvalues appear ($T=1.60\times 10^{-1}$ in Fig. \ref{TVI_Sx_order_H040_EVD}).
In the low-temperature region, the quantum effect due to the $S^zS^z$ term becomes dominant, where dense small loops of $S^x=\pm 1/2$ implicitly represent the classical ordered state in the $S^z$ direction.
Thus, number of small loops in WLs increases even for the ordered groundstate, and the snapshot spectra show quite different behaviors from those in the $S^z$ basis in the previous section.
At the lowest temperature ($T=1.25\times10^{-3}$), $P(\omega)$ actually has a dome-like shape with no population at $\omega=0$, in addition to the $\delta$-function-like peak of the maximum eigenvalue.
The reason for the maximum-eigenvalue peak is that population of the $S^x=+1/2$ loops is explicitly biased by the finite $\Gamma$ in the $S^x$ basis.

\begin{figure}[tb]
\centering\includegraphics[width=0.8\linewidth]{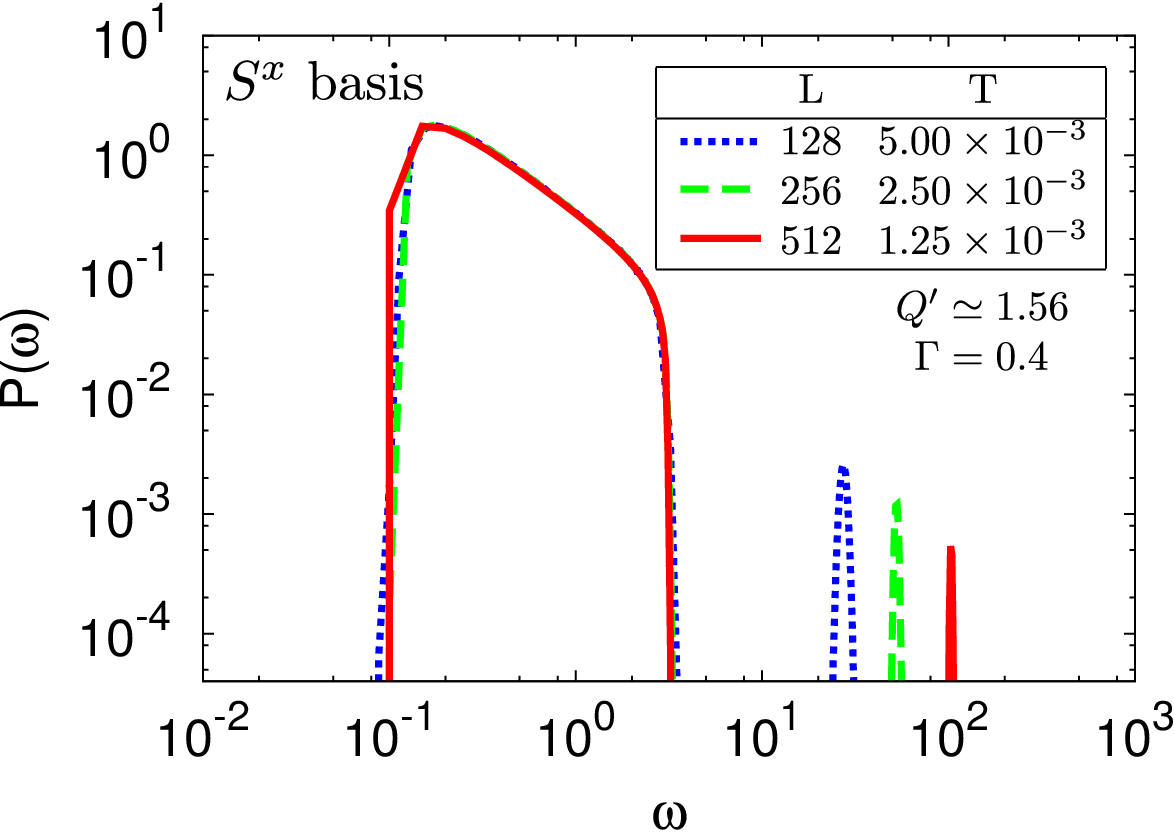}
\caption{
(Color online)
Eigenvalue distribution $P(\omega)$ in the $S^x$ basis for $\Gamma=0.4$ in the low-temperature regime.
$P(\omega)$ for various $L$ and $T$ with the fixed effective aspect ratio $Q'=\beta J/L=25/16 \simeq 1.56$ collapse onto the dome-like curve characterized by the aspect ratio $Q'$, except for the maximum eigenvalue.
}\label{TVI_Sx_order_H040_fixed_EVD}
\end{figure}

The situation of the dome-like structure of $P(\omega)$ seems similar to that for the $S^z$-diagonal basis in the disorder phase (see Fig. \ref{TVI_Sz_disorder_H200_EVD}).
Here, we investigate the low-temperature property of $P(\omega)$ for the $S^x$ basis in details.
An essential point is that, in the $S^x$ basis, both of the spin-spin interaction and the quantum fluctuation are represented by the $S^zS^z$ term in the Hamiltonian.
This implies that the effective system size in the imaginary-time direction is $\beta J$, in contrast to $\beta \Gamma$ for the case of the $S^z$ basis in Sec. \ref{TVI_Sz_disorder}.
We thus adopt
\begin{equation}
Q' = \frac{\beta J}{L},
\label{aspectratio_Q2}
\end{equation}
as an effective aspect ratio, instead of $Q$.
In Fig. \ref{TVI_Sx_order_H040_fixed_EVD}, we then show $P(\omega)$ for various $T$ and $L$ with a fixed $Q'(=25/16 \simeq 1.56)$ in sufficiently low temperatures.
We can confirm that the distribution functions collapse to the dome-like shape specified by $Q'$, where the contribution of the maximum-eigenvalue peak is negligible (${\cal O}(1/L)$ to the dominant part).

\begin{figure}[tb]
\centering\includegraphics[width=0.8\linewidth]{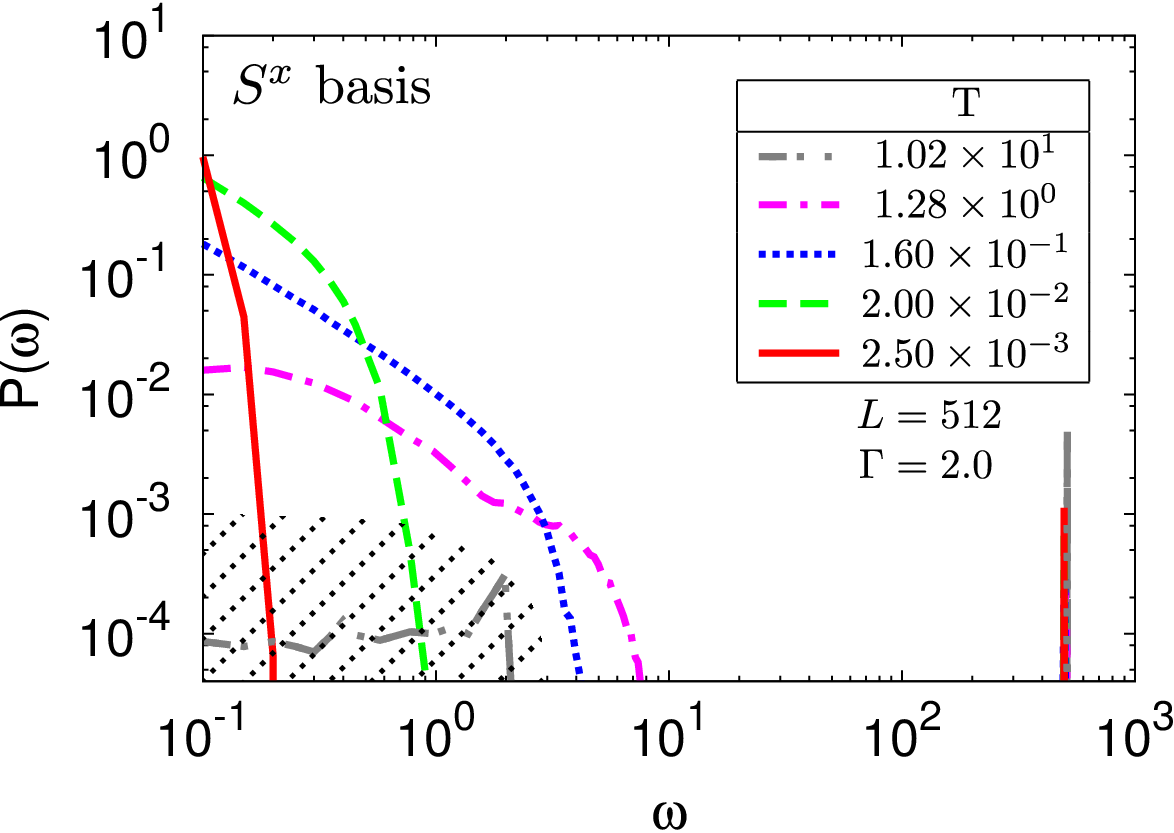}
\caption{
Temperature dependence of the eigenvalue distribution $P(\omega)$ in the $S^x$ basis for the transverse-field Ising chain with $\Gamma=2.0$ and $L=512$.
The statistical error is negligible except for the shaded region of small $\omega$.
}\label{TVI_Sx_disorder_H200_EVD}
\end{figure}

In Fig. \ref{TVI_Sx_disorder_H200_EVD}, we next show $P(\omega)$ in the $S^x$ basis for the disordered phase ($\Gamma=2.0$).
At high-temperature, the eigenvalue spectrum has the isolated peak of the maximum eigenvalue and macroscopic number of eigenvalues condensate at $\omega=0$ (Note that $\omega=0$ is out of range due to the log scale plot).
In Fig. \ref{TVI_Sx_disorder_H200_EVD}, we can see that this double peak structure is basically retained in all temperatures, where dominant number of WLs has $S^x=1/2$ due to the large $\Gamma$.
As $T$ decreases, thus, the degenerating eigenvalues at $\omega=0$ are weakly disturbed by the quantum fluctuation.
We can actually observe the narrow distribution in the vicinity of $\omega=0$ below the crossover temperature ($T\sim J$), implying the robust $\omega=0$ peak.
Note that this crossover temperature is almost independent of $\Gamma$, since the quantum fluctuation is governed the energy scale $J(=1)$ rather than $\Gamma$.

\begin{figure}[tb]
\centering\includegraphics[width=0.8\linewidth]{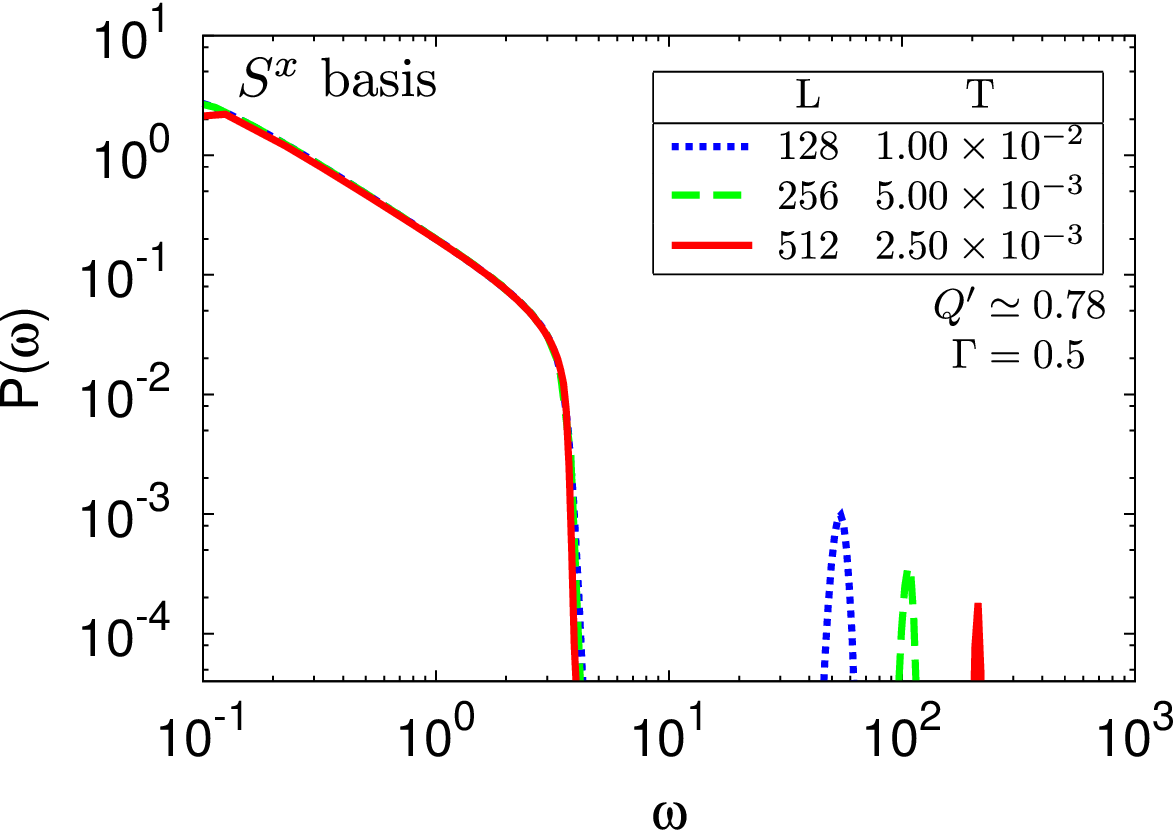}
\caption{
(Color online)
Eigenvalue distribution $P(\omega)$ in the $S^x$ basis at the critical point ($\Gamma=0.5$) in low temperatures.
$P(\omega)$ collapses into the curve characterized by the aspect ratio $Q' =25/32 \simeq 0.78$.
}\label{TVI_Sx_critical_H0500_EVD}
\end{figure}

In Fig. \ref{TVI_Sx_critical_H0500_EVD}, we finally present the eigenvalue distributions for the sufficiently low temperature at the critical point ($\Gamma=0.5$) with the fixed $Q'=25/31(\simeq 0.78)$.
A significant difference from the $S^z$-basis result in Fig. \ref{TVI_Sz_critical_H0500_EVD} is that the continuous distribution abruptly drops at $\omega\simeq 3.0$, where no power-law behavior is observed and the $\delta$-function-like peak of the maximum eigenvalue remains.

In order to analyze this feature in Fig. \ref{TVI_Sx_critical_H0500_EVD}, let us recall that the SDM at the critical point can be related with the correlation function matrix through the self-averaging effect.
For the case of the $S^x$-diagonal basis, we may assume
\begin{equation}
\rho^x(n, m)\sim \langle S_n^x S_m^x \rangle \sim {\rm const} \times|n-m|^{-\eta_x} + \langle S^x \rangle^2,
\label{sdm_x}
\end{equation}
where a finite $\langle S^x \rangle$ is induced by the finite $\Gamma$.
Since the SDM of Eq. (\ref{sdm_x}) contains $\langle S^x \rangle^2$ in its matrix elements as a background, the maximum eigenvalue turns out to be $\omega \simeq L \langle S^x \rangle^2 $, which can be actually verified in the numerical results in Fig. \ref{TVI_Sx_critical_H0500_EVD}.
Also, we have checked the area of the maximum-eigenvalue peak being $1/L$, supporting that the $\delta$-function-like peak solely consists of the maximum eigenvalue.
For the sudden drop of $P(\omega)$ at $\omega \simeq 3.0$ in Fig. \ref{TVI_Sx_critical_H0500_EVD}, an important point is $\eta_x=2$ for the transverse-field Ising chain\cite{pfeuty1970}.
In contrast to the $S^z$ basis, the exponent $\eta_x$ for Eq. (\ref{sdm_x}) is  $\eta_x > 1$, for which Eq. (\ref{disp_omega}) diverges due to a singularity at $r = 0$.
This implies that $P(\omega)$ has the singular edge of the continuous distribution, which is consistent with the distribution edge at $\omega \simeq 3.0$ in Fig. \ref{TVI_Sx_critical_H0500_EVD}.

In the remaining part of this section, we would like to discuss the snapshot spectrum from the viewpoint of the duality relation.
As well known, the transverse-field Ising chain has the Kramers-Wannier duality transformation\cite{duality}, which exchanges roles of the $S^zS^z$ term and the $S^x$ term in the Hamiltonian through $\tilde{\Gamma}= J/2$ and $\tilde{J} =  2\Gamma$\cite{duality2}.
We can also apply the duality transformation in the WL snapshot level, which is summarized as follows.
For the case of converting a snapshot in the $S^z$ basis to that in the $S^x$-diagonal basis, 
(a) If two adjacent spins have the same direction, put a ${S}^x=+1/2$ spin on the dual lattice point, and 
(b) If two adjacent spins have the opposite directions, put a ${S}^x=-1/2$ spin on the dual lattice point.
Then, we can map the WLs representing the classical order in the $S^z$ basis for $\Gamma/J <0.5$ into macroscopic-size WL loops  in the $S^x$ basis for $\tilde{\Gamma}/\tilde{J} > 0.5$.
Thus, the situation of the large loops is similar to the ordered spins in the $S^z$ basis.
Similarly, WLs of frequently flipping spins for $\Gamma/J>0.5$ in the $S^z$ basis are mapped to dense small loops in the $S^x$ basis for $\tilde{\Gamma}/\tilde{J}<0.5$.
Thus, we can qualitatively deduce the properties of $P(\omega)$ in the $S^x$ basis as the duality of the distribution in the $S^z$ basis.
Near $\Gamma/J=0.5$, however, the snapshots in the two basis exhibit quite distinct appearances in spite of the self-dual point, reflecting the differences between $\langle S^z_n S^z_m \rangle$ and $\langle S^x_n S^x_m \rangle$.

\section{XXZ chain}
In this section, we investigate the snapshot spectrum for the $S = 1/2$ XXZ chain, which is defined as
\begin{equation}
\mathcal{H} = J\sum_{n}^{L} (\hat{S}_n^x\hat{S}_{n+1}^x + \hat{S}_n^y\hat{S}_{n+1}^y + \Delta\hat{S}_n^z\hat{S}_{n+1}^z).
\end{equation}
We fix $J=1$ for simplicity.
The QMC simulation for the XXZ chain is also based on the loop algorithm in Ref. \cite{kawashima2004recent}.
From the viewpoint of the snapshot spectrum, an interesting point on the XXZ chain is that the quantum fluctuation is introduced by the XY term, in contrast to the transverse-field Ising chain where the transverse magnetic field locally induces the quantum fluctuation.

Before presenting numerical results, we briefly summarize the groundstate properties of the XXZ chain essential for analyses of the snapshot spectrum.
For $\Delta > 1$, the groundstate has the Neel order with the excitation gap.
In the $-1 < \Delta \le 1$ region, the groundstate is critical, where the correlation functions obey the power-law behaviors\cite{luther1975calculation},
\begin{gather}
\langle S_i^xS_{i+n}^x \rangle \sim n^{-\eta} \label{xxz_corr_x}\\
\langle S_i^zS_{i+n}^z \rangle \sim n^{-1/\eta} \label{xxz_corr_z}.
\end{gather}
The exact value of $\eta$ is obtained as
$
\eta = 1-\frac{\arccos\Delta}{\pi}
$
with the Bethe ansatz\cite{Bogoliubov}.
In the following, we perform QMC simulations to generate WL snapshots in the $S^z$ and $S^x$ diagonal basis for $\Delta > -1$, and then discuss the temperature dependence of $P(\omega)$ in details.

\subsection{Ising-like region ($\Delta = 2.0$)}
\begin{figure}[tb]
\centering\includegraphics[width=0.8\linewidth]{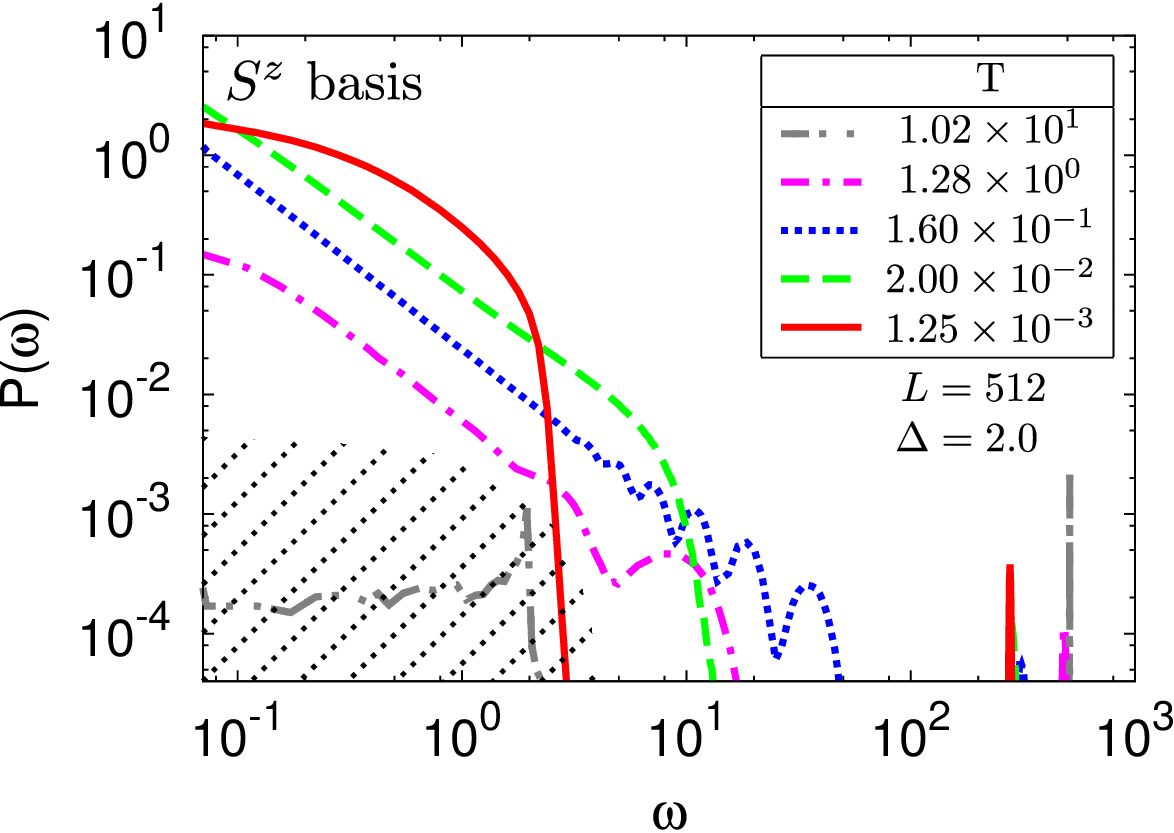}
\caption{
(Color online)
Temperature dependence of the eigenvalue distribution $P(\omega)$ in the $S^z$ basis for the Ising-like XXZ chain with $\Delta = 2.0$ and $L=512$.
The statistical error except for the shaded region is negligible.
}\label{XXZ_Delta20_Sz_EVD}
\end{figure}

We consider the snapshot spectrum in the $S^z$-diagonal basis for the Ising-like XXZ chain, which has the classical groundstate order.
Figure \ref{XXZ_Delta20_Sz_EVD} shows temperature dependence of $P(\omega)$ for $\Delta = 2.0$ with the system size $L=512$, which is sufficiently longer than the groundstate correlation length.
In the high-temperature limit, $P(\omega)$ has two peaks at the maximum eigenvalue $\omega \simeq L(=512)$ and $\omega = 0$, as in the case of the transverse-field Ising chain.
Here, note that the peak of $\omega=0$ is out of the range in Fig. \ref{XXZ_Delta20_Sz_EVD}, as well.
As temperature decreases, the quantum fluctuation due to the XY term disturbs the classical WL configurations.
We then observe that, around the crossover temperature $T\sim J/\Delta = 0.5$, $P(\omega)$ in the low-$\omega$ region develops to dominant distributions having a multi-peak structure.
As temperature decreases further, the multi-peak structure is smeared out and finally disappears at the lowest temperature ($T = 1.25 \times 10^{-3}$).
However, the $\delta$-function-like peak of the maximum eigenvalue remains around $\omega \simeq 300$, reflecting the existence of the Neel order in the groundstate.
The above behavior is qualitatively reminiscent of the spectrum for the transverse-field Ising chain in the $S^z$ basis, where the classical order also exists at the groundstate.

\begin{figure}[tb]
\centering\includegraphics[width=0.8\linewidth]{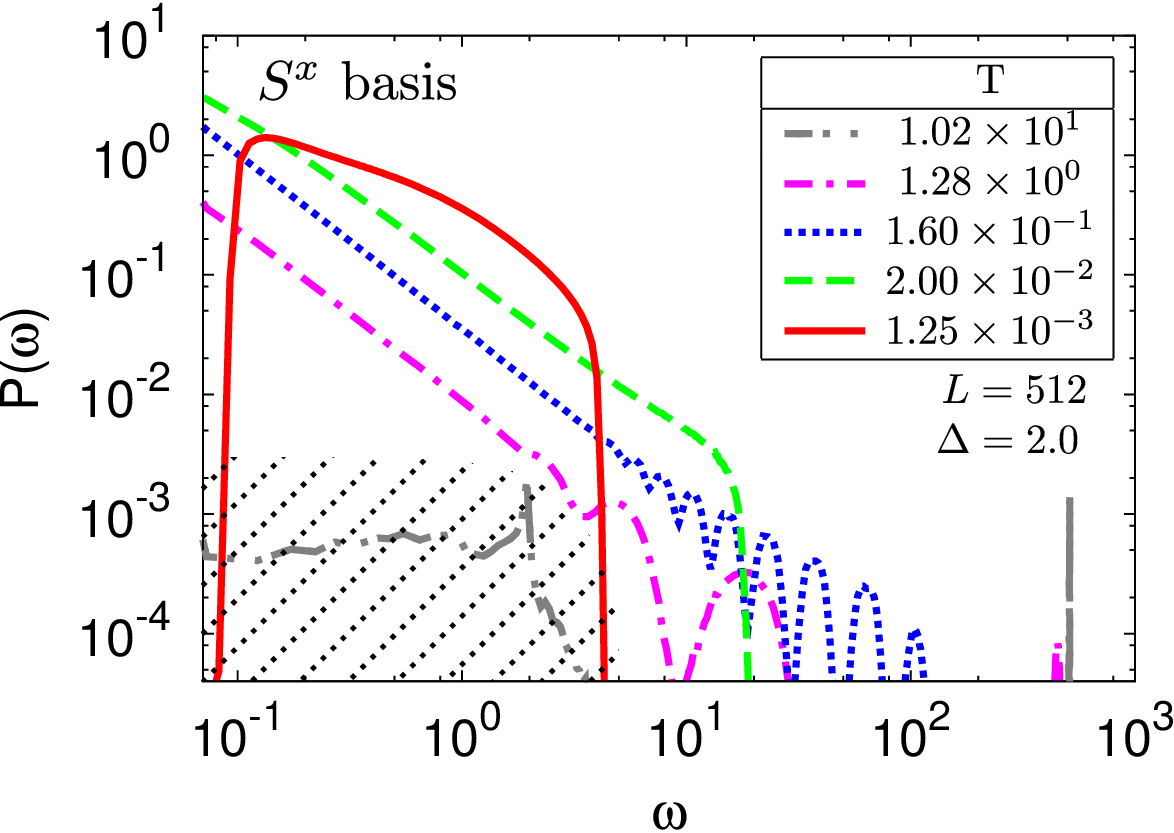}
\caption{
(Color online)
Temperature dependence of the eigenvalue distribution $P(\omega)$ in the $S^x$ basis for $\Delta = 2.0$ and $L=512$.
The statistical error except for the shaded region is negligible.
}\label{XXZ_Delta20_Sx_EVD}
\end{figure}

We turn to the snapshot spectrum in the $S^x$-diagonal basis for the same parameters, $\Delta=2.0$ and $L=512$, which is presented in Fig. \ref{XXZ_Delta20_Sx_EVD}.
As temperature decreases, the broad distribution appears in the low-$\omega$ region, and the $\delta$-function-like peak of the maximum eigenvalue shifts to the small-$\omega$ side.
As temperature decreases further, the maximum-eigenvalue peak is absorbed into the low-$\omega$ distribution and finally $P(\omega)$ converges to a dome-like shape in the low-temperature limit ($T=1.25\times 10^{-3}$).
In order to analyze this dome-like shape in the low-temperature limit, we fix an effective aspect ratio, $\frac{\beta J\Delta }{L}$, since the competition between the classical order and the quantum fluctuation is characterized by the anisotropy coefficient $\Delta$ of the XXZ chain.
Indeed, we have confirmed that $P(\omega)$ for $\Delta>4$ converges to the dome-like curve specified by the fixed aspect ratio of $\frac{\beta J \Delta }{L}$.
(The numerical result for this is not presented here.)

\subsection{Critical regime}
In the $|\Delta| < 1$ region, the groundstate of the XXZ chain is critical, where the power-law behavior of $P(\omega$) is also expected in sufficiently low temperatures.
In the critical regime,  we should recall that the SDM for the transverse-field Ising chain can be related to the corresponding correlation functions through Eq. (\ref{sdm_z}), where the exponent $\alpha$ for the power-law behavior of $P(\omega)$ is given by Eq. (\ref{exponent_alpha}).

\begin{figure}[tb]
\centering\includegraphics[width=0.8\linewidth]{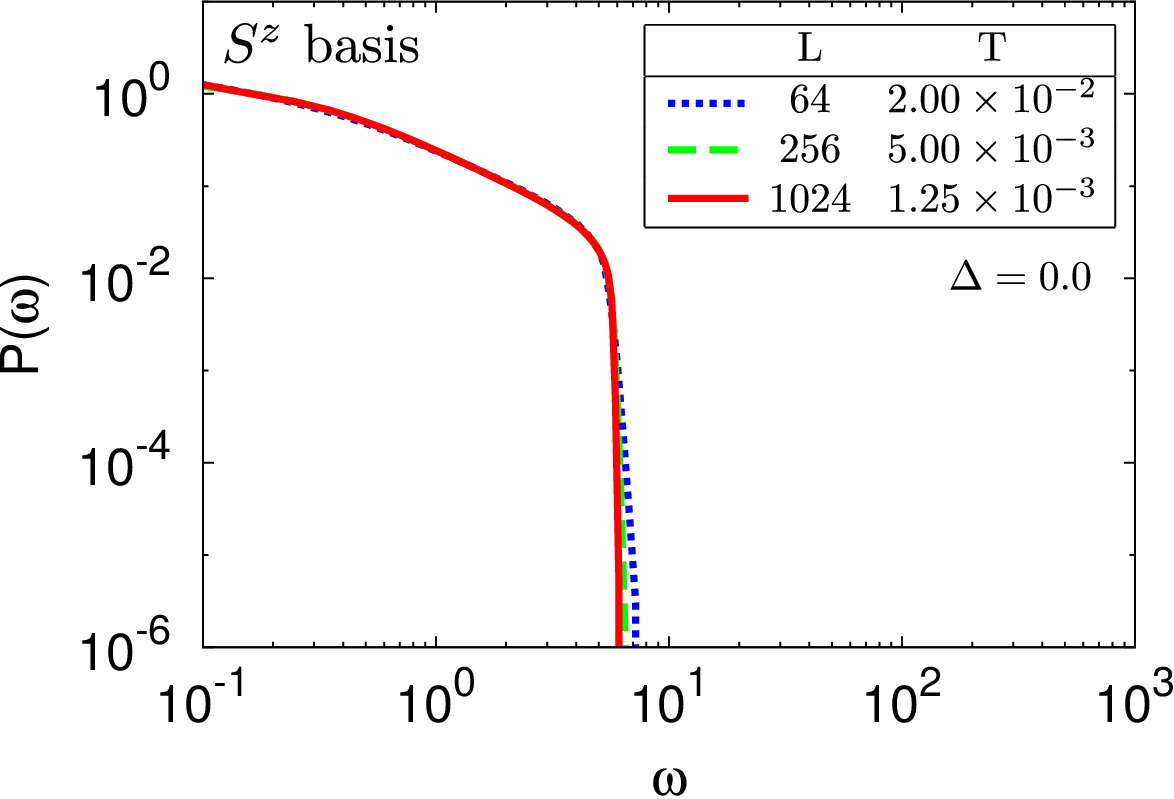}
\caption{
(Color online)
Low-temperature behavior of the eigenvalue distribution $P(\omega)$ in the $S^z$-diagonal basis for the XY chain ($\Delta = 0.0$).
$P(\omega)$ has a singular distribution edge at $\omega \simeq 5.0$.
}\label{XXZ_Delta05_Sz_critical_EVD}
\end{figure}

We first consider the snapshot spectrum in the $S^z$ basis, where the behavior of $\langle S^z_i S^z_{i+n}\rangle$ is important.
In the region of $|\Delta|< 1$, the exponent of Eq. (\ref{xxz_corr_z}) always satisfies $1/\eta >1$.
Then, Eq. (\ref{disp_omega}) for the SDM in the $S^z$ basis diverges at $r=0$, suggesting that $P(\omega)$ has a sharp edge of the distribution.
Figure \ref{XXZ_Delta05_Sz_critical_EVD} shows $P(\omega)$ for the XY chain ($\Delta=0.0$) in the $S^z$ basis in low temperatures.
Note $1/\eta = 2$ for the XY chain.
In the figure, the singular edge of $P(\omega)$ is actually observed at $\omega \simeq 5.0$, which is consistent with the theoretical analysis based on Eq. (\ref{disp_omega}).
Also,  the singular edge of the distribution is confirmed for $\Delta= \pm 0.5$.

\begin{figure}[tb]
\centering\includegraphics[width=0.8\linewidth]{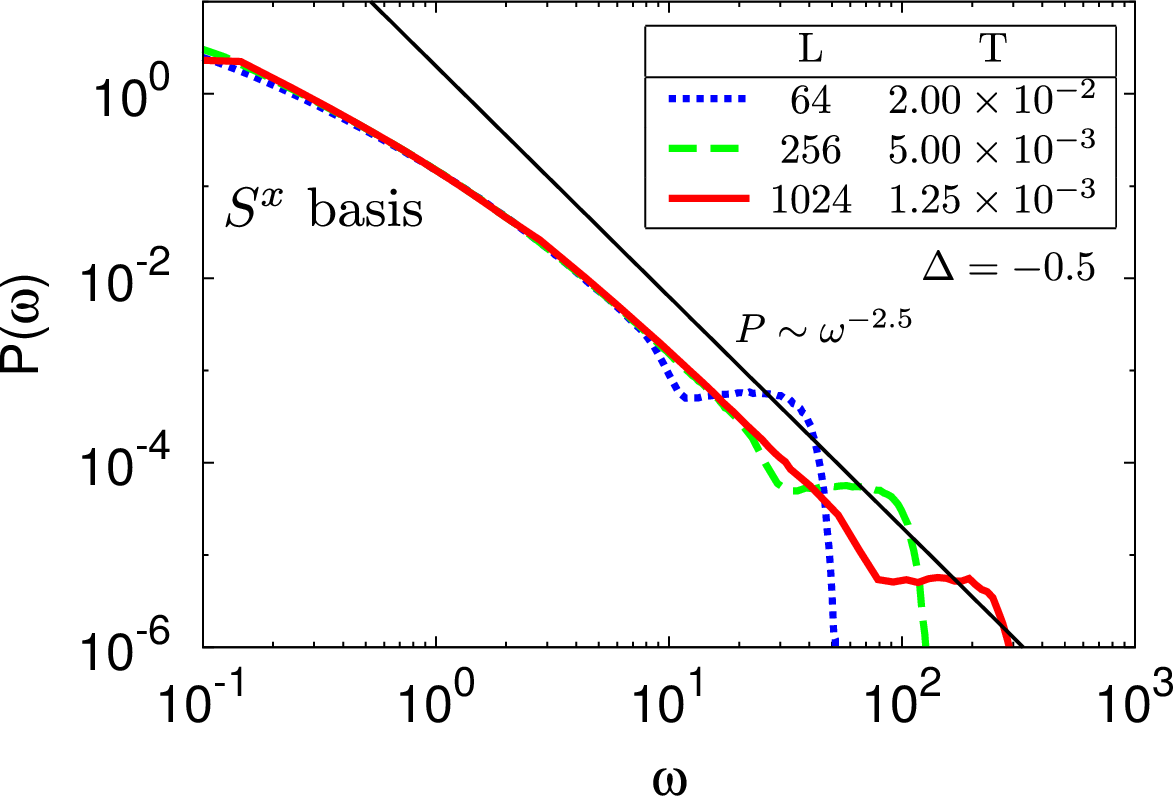}
\caption{
(Color online)
Log-log plot of the eigenvalue distribution $P(\omega)$ in the $S^x$ basis for the XXZ chain with $\Delta=-0.5$.
The inverse temperature and the system size is scaled with the fixed aspect ratio $\beta /L(= 25/32)\simeq 0.78$.
The solid line indicates a slop with the exponent $\alpha=5/2$ for comparison.
}\label{XXZ_Delta05_Sx_critical_EVD}
\end{figure}

We next discuss the critical behavior of the snapshot spectrum in the $S^x$-diagonal basis, which can be related to $\langle S_i^xS_{i+n}^x \rangle$.
In Fig. \ref{XXZ_Delta05_Sx_critical_EVD}, we present $P(\omega)$ for $\Delta=-0.5$, where the critical exponent is given by $\eta = 1/3$.
Since the dominant quantum fluctuation in the $S^x$-diagonal basis for $|\Delta|<1$ is governed by the $S_{i}^y S_{i+1}^y$ term in the Hamiltonian, we should scale $\beta$ and $L$ with the fixed aspect ratio $\beta/L$.
In Fig. \ref{XXZ_Delta05_Sx_critical_EVD}, $P(\omega)$ extends to the larger-$\omega$ side with the power-law behavior, as $\beta$ increases.
Then, we can verify that the slope of $P(\omega)$ is consistent with $\alpha=5/2$, which is obtained through Eq. (\ref{exponent_alpha}) with $\eta=1/3$.

Here, we would like to comment on the finite-size effect on $\alpha$.
Although the numerical estimation of $\alpha$ is consistent with the exact value $\alpha=5/2$ for $\Delta =-0.5$, we note that, as $\Delta$ increases, the region where the correct power-law behavior is observed shifts toward a larger-$\omega$ region, for which we need a larger system size (or equivalently a larger $\beta$).
For the XY chain ($\Delta=0$), indeed, we have diagonalized the exact correlation function matrix\cite{Lieb_Schultz_Mattice} to check the consistency with the snapshot spectrum.
We then find that the finite-size effect for $\Delta=0$ becomes more significant than that for $\Delta=-0.5$.
A reason for this significant size effect may be attributed to the fact that, as $\Delta$ increases, $\eta$ also approaches $\eta= 1$ at the Heisenberg point where Eq. (\ref{exponent_alpha}) diverges.

\section{Summary}
In this paper, we have introduced the snapshot density matrix (SDM) for the world-line (WL) quantum Monte Carlo (QMC) and investigated its eigenvalue spectrum for the transverse-field Ising chain and the XXZ chain in details.
We define the SDM as Eq. (\ref{SDMc}) by tracing out the continuous-imaginary-time index of WL snapshots.
For the transverse-field Ising chain, we have actually calculated the distribution function $P(\omega)$ for the snapshot spectrum, and revealed its fundamental properties: 
(i) In the ordered phase, the isolated peak corresponding to the maximum eigenvalue always appears, indicating the classical nature of WL configurations.
(ii) In the disordered phase, $P(\omega)$ converges to the dome-like distribution specified by the appropriate aspect ratio defined by Eq. (\ref{aspectratioQ}).
(iii) At the critical point, the power-law distribution with the exponent of Eq. (\ref{exponent_alpha}), which can be related to the correlation function, is observed in the low-temperature limit with the fixed aspect ratio.

In addition, we clarified properties of the snapshot spectrum for WLs in the $S^x$-diagonal basis, where the quantum fluctuation is attributed to the $S_{i}^zS_{i+1}^z$ term.
Using another aspect ratio of Eq. (\ref{aspectratio_Q2}), we have extracted essential features of the distributions in the low-temperature limit.
We can also explained the properties of the spectrum in the $S^x$-diagonal basis from the viewpoint of the Kramers-Wannier duality.
For the XXZ chain, we have investigated the snapshot spectrum as well. 
Particularly in the critical region of $|\Delta| <1$, we have obtained the proper power-law distribution for the $S^x$ basis representation, which is consistent with the analysis based on the correlation function.

The snapshot spectrum represents weight for hierarchical decomposition of winding/entangling WLs, which provides a novel point of view for understanding the role of the quantum fluctuation in the WL level.
Actually, we have revealed a series of interesting properties of the snapshot spectrum for the WL QMC, reflecting the groundstate phases.
However, how the snapshot spectrum can be connected to the quantum entanglement for the groundstate is not clear in the present level.
In order to answer our original motivation, further researches are needed.

\begin{acknowledgements}
This work was supported in part by Grants-in-Aid No. 26400387, 16J02724 and 17H02931 from Japan Society of the Promotion of Science.
\end{acknowledgements}

\appendix

\section{World-line quantum Monte Carlo}
\label{appendix0}
\begin{figure}[tb]
\centering\includegraphics[width=0.7\linewidth]{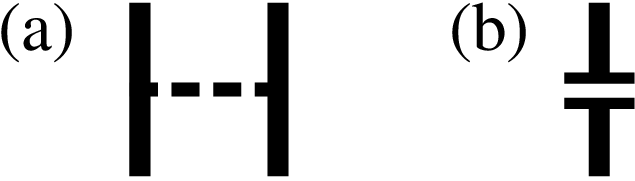}
\caption{
(Color online)
Graph elements for the transverse-field Ising model.
(a) ``Bind graph" connects two adjacent WLs having a parallel spin direction.
(b) ``Cut graph" is inserted at a possible location of spin flipping (kink) in WLs. 
}\label{fig_A1}
\end{figure}

\begin{figure}[tb]
\centering\includegraphics[width=0.8\linewidth]{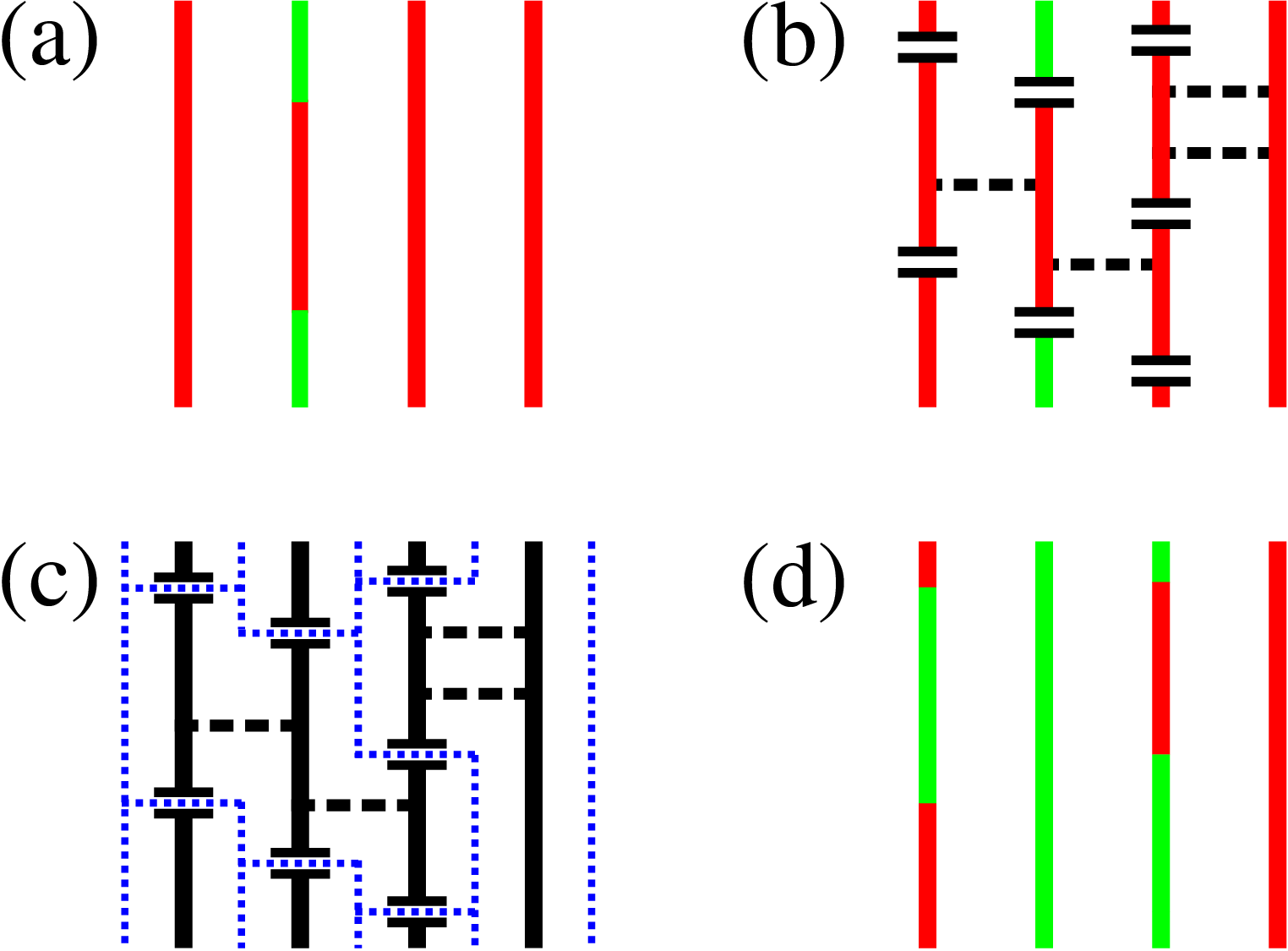}
\caption{
(Color online)
Update scheme of the WL configuration in the loop algorithm.
Red and green lines respectively represent WLs of up and down spins.
(a) A given WL configuration of the transverse-field Ising model.
(b) The graphs assigned for the WL configuration.
(c) Cluster analysis for WLs connected by bind graphs.
The WLs surrounded by blue dotted lines form WL clusters.
(d) Up or down spins are randomly assigned for the clusters.
}\label{fig_A2}
\end{figure}

For the ferromagnetic transverse-field Ising chain in the $S^z$ basis representation, we briefly explain how matrix elements of the Hamiltonian can be related with structures of WLs generated by WL QMC.
In WL QMC, we represent the partition function of the chain as a trace of weighted WLs of spins with the continuous imaginary time index on the basis of the path integral representation,  and then perform sampling of them with a cluster-like update scheme.
For updating of a WL snapshot in the loop-type algorithm\cite{kawashima2004recent}, we consider two types of graph elements for WLs:
 ``bind graph'' and ``cut graph'' (See Fig. \ref{fig_A1}), which are respectively associated with the  $S^zS^z$ interaction term and the $S^x$ term.
For a given WL configuration [Fig. \ref{fig_A2} (a)], the bind graph is placed for two adjacent WLs carrying the same spin index with a density $ J/2$ per unit length in the imaginary-time direction.
Note that the total length of the imaginary-time direction is given by the inverse temperature $\beta$.
Also, the cut graph, which represents spin flipping due to $S^x$, is assigned for WLs with a density $\Gamma/2$, in addition at the kink places already included in the given WLs [Fig. \ref{fig_A2} (b)].

After the allocation of the graph elements, we perform cluster analysis for WLs connected by the bind graphs [Fig. \ref{fig_A2} (c)].
According to the $Z_2$ symmetry of the transverse Ising chain,  then, we randomly assign spin directions to the WL clusters [Fig. \ref{fig_A2} (d)], and obtain a new WL configuration.
Note that the above sampling of WLs corresponds to that based on the Suzuki-Trotter decomposition in the infinite Trotter number limit.
Thus, the distribution of the graph elements in WLs and the resulting WL shapes reflect relative relevances of the classical $S^zS^z$ interaction and the quantum fluctuation due to the $S^x$ term in the thermal equilibrium.
Note that WLs for the transverse-field Ising model do not necessarily form closed loops, where the on-site $S^x$ term trivially breaks the $S^z$ conservation. 
On the other hand, WLs for the XXZ chain always draw closed loops, where the exchange-type interaction include not a cut graph but exchange graphs of WLs.

\section{Hierarchical decomposition  of a world-line snapshot}
\label{appendix1}
\begin{figure}[tb]
\centering\includegraphics[width=1.0\linewidth]{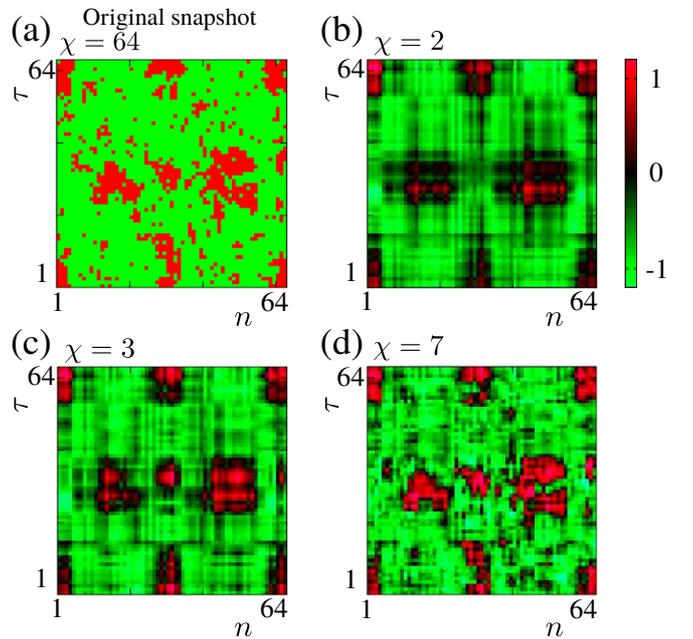}
\caption{
(Color online)
Snapshot matrix for the transverse-field Ising chain at $\Gamma = 0.5$ and $T=1/64$ with $L=64$ and $N_{\beta} =64$, where the horizontal and vertical axes respectively represent the spatial and imaginary-time directions.
(a) Original snapshot matrix $M^z(n,\tau_i)$.
Red and green pixels represent up ($+1$) and down ($-1$) spins respectively.
(b)  $M_{\chi=2}^z(n,\tau_i)$ reproduces the global structure of WLs.
(c)  Intermediate-size spin clusters appear in $M_{\chi=3}^z(n,\tau_i)$.
(d)  Most of fine structures can be seen in $M_{\chi=7}^z(n,\tau_i)$.
}\label{hierarchical_structure}
\end{figure}

In this appendix, we demonstrate the hierarchical structure of the snapshot spectrum for a typical WL snapshot.
As an example, we consider a WL snapshot of the transverse-field Ising chain at $\Gamma = 0.5$ with the system size $L=64$ and the trotter number $N_\beta =64$.
Figure \ref{hierarchical_structure}(a) shows a typical snapshot generated by QMC at the temperature $T=1/64$, where $M^z = \pm 1$ pieces of the discretized WLs are drawn as red/green pixels. 
For this snapshot, we directly perform the SVD of Eq. (\ref{svd}) to have $\omega_l$ and the corresponding singular vectors $U_l$ and $V_l$. 
Then, a key point is that $\omega_l$ can be viewed as weight amounting to importance of the singular vectors $U_l(n) V_l(\tau_i)$ in the entire WL snapshot;
For the largest eigenvalue $\omega_1$,  $U_1$ and $V_1$ represent the most global structure of the WL configuration. 
As $\omega_l$ decreases, the corresponding $U_l$ and $V_l$ systematically reproduce fine structures of the WLs.

We define the snapshot matrix reconstructed up to a cutoff dimension $\chi ( \le 64$) as
\begin{equation}
M_{\chi}^z(n, \tau_i) = \sum_{l=1}^{\chi} U_l(n)\sqrt{\omega_l} V_l(\tau_i),
\end{equation}
which gives an approximated version of the original snapshot.
In  Fig. \ref{hierarchical_structure}(b), we can see that $M_{\chi=2}^z$ well reproduces the global structure of the original snapshot, even though it contains only two singular values.
As shown in Fig.  \ref{hierarchical_structure}(c),  the intermediate-size spin clusters gradually emerge with increasing $\chi$.
In  Fig. \ref{hierarchical_structure}(d), moreover,  we can see that the most of fine structures of spin clusters are well approximated in  $M^z_{\chi=7}$.
Therefore, the distribution of $\omega_l$ particularly in the larger-$\omega$ region provides a novel characterization of the system behind the hierarchical structure of the WL configurations, complementally to the bulk expectation values of physical quantities.

\end{document}